# Evolution of the 2012 July 12 CME from the Sun to the Earth: Data-Constrained Three-Dimensional MHD Simulations

Fang Shen[1], Chenglong Shen[2], Jie Zhang[3], Phillip Hess[3], Yuming Wang[2], Xueshang Feng[1], Hongze Cheng[2] and Yi Yang[1]

[1]SIGMA Weather Group, State Key Laboratory of Space Weather, Center for Space Science and Applied Research, Chinese Academy of Sciences, Beijing 100190, China

[2]School of Earth & Space Sciences, University of Science & Technology of China, Hefei, Anhui 230026, China

[3]School of Physics, Astronomy and Computational Sciences, George Mason University, 4400 University Drive, MSN 6A2, Fairfax, VA 22030, USA

Corresponding author: Fang Shen, SIGMA Weather Group, State Key Laboratory of Space Weather, Center for Space Science and Applied Research, Chinese Academy of Sciences, Beijing 100190, China. (fshen@spaceweather.ac.cn)

# Abstract

The dynamic process of coronal mass ejections (CMEs) in the heliosphere provides us the key information for evaluating CMEs' geo-effectiveness and improving the accurate prediction of CME-induced Shock Arrival Time (SAT) at the Earth. We present a data-constrained three-dimensional (3D) magnetohydrodynamic (MHD) simulation of the evolution of the CME in a realistic ambient solar wind for the July 12-16, 2012 event by using the 3D COIN-TVD MHD code. A detailed comparison of the kinematic evolution of the CME between the observations and the simulation is carried out, including the usage of the time-elongation maps from the perspectives of both Stereo A and Stereo B. In this case study, we find that our 3D COIN-TVD MHD model, with the magnetized plasma blob as the driver, is able to re-produce relatively well the real 3D nature of the CME in morphology and their evolution from the Sun to Earth. The simulation also provides a relatively satisfactory comparison with the in-situ plasma data from the Wind spacecraft.

1. Introduction

Coronal mass ejections (CMEs) are a large-scale eruption of magnetized plasma from the Sun's corona and subsequently propagate into interplanetary space. They are the main drivers of space weather near the Earth, accounting for about 85% of intense geomagnetic storms [Zhang et al., 2007], especially when they contain organized southward-directed magnetic fields. However, not all of CMEs originating from the vicinity of solar center can encounter the Earth [e.g., Wang et al., 2002; Yermolaev and Yermolaev, 2006; Shen et al., 2014 and references therein]. Therefore, it is of great importance to understand how they propagate and evolve in interplanetary space, and how their properties as observed at 1 AU are related to the properties observed near the Sun [e.g., Lugaz et al., 2011].

In recent years, numerical simulations of interplanetary coronal mass ejections (ICMEs) have become one of the primary tools to investigate the propagation of ICMEs and their interaction with the interplanetary medium. The in-depth study of the evolution of CMEs in the heliosphere heavily depends upon numerical models. There have been Sun-to-Earth numerical simulations of real events [e.g., Chané et al., 2008; Lugaz et al., 2007, 2011; Tóth et al., 2007; and Shen et al., 2007, 2011a, 2011b]. Particular attention has been given to the numerical modeling of CMEs at or near solar minimum, especially the May 12, 1997 CME [Odstrcil et al., 2004; Wu et al., 2007; Cohen et al., 2008a; Titov et al., 2008; Zhou et al., 2008]. One reason for such choice is that the ambient solar wind is believed to be simpler and steadier during solar minimum, thus easier to modeling, than that during the solar maximum. Therefore, solar minimum is thought to be the perfect period to study the evolution of CMEs. Meanwhile, there still exist a few numerical works of CME events at or near solar maximum, e.g., the April 4, 2000 CME [Chané et al., 2008; Shen et al., 2011a].

However, not all of these previous works have been totally successful in reproducing the observed transit time and the measured plasma properties at 1 AU, especially for the CMEs near the maximum. One of the possible reasons is that some parameters, particularly the initial speeds, of CMEs for those events are not well constrained, because Earth-directed CMEs appearing halo suffer from the projection effect and there were no direct observations then between 32 Rs and Earth. With the launch of the twin Solar Terrestrial Relations Observatory (STEREO) spacecraft in 2006 [Kaiser, et al., 2008], CMEs can be imaged continuously from the solar surface to 1AU with coronagraphic and heliospheric imagers onboard STEREO [Howard, et al., 2008]. This provides the opportunity to directly compare the simulation results with the observations continuously in time and space. We believe that constructing a data-constrained numerical model is necessary to reproduce the measured plasma properties at 1 AU, in which both the imaging data and the *in-situ* data should be used to constrain the initial parameters of the CME, including propagation direction, speed, density, temperature and magnetic field.

In this article, we study the kinetic evolution of the July 12-16, 2012 CME event. The event was recently studied by several authors using observational or theoretical method [Moestl et al., 2014; Cheng et al., 2014; Dudík et al., 2014; Hess and Zhang, 2014]. Thus, it is intriguing to study the July 12-16, 2012 CME event using the data-constrained 3D MHD numerical method and compare the simulated Sun-to-Earth evolution results with actual observations in 3D space and continuously in time. The organization of the paper is as follows. We describe the observations and numerical models in Section 2. Details of the kinematic evolution of the CME in interplanetary space is discussed in Section 3. This section also includes the synthetic STEREO-like line-of-sight images, and the comparison of the time-elongation maps (J-maps) between synthetic results and white-light observations. The comparison with the in-situ data at 1AU is also explored. In the last section, a summary and discussion is given.

**2. Observations and methods**

The event on July 12-16, 2012 is a fast, Earth-directed CME occurring at the solar maximum in the 24th solar cycle and belonging to Carrington Rotation (CR) 2125, with an initial speed of 1531 km/s at 2 Rs and initial direction of S09W01 [Hess and Zhang, 2014]. Over July 14-15, 2012 and through the early time of July 16, the Earth experienced a strong geomagnetic storm with peak Dst of -127 nT, which created an aurora visible at lower latitudes of the Earth. This event was very well observed and tracked by the imaging instruments on STEREO due to the optimal propagation direction of the CME and the viewing angles of STEREO A and B.

**2.1 The Solar Eruption of 2012 July 12**

The CME of interest originated from AR 11520. The active region first appeared on the solar disk from the eastern limb on about 2012 July 5 and rotated beyond the western limb on about 2012 July 19. During this period, this active region produced seven M class flares and one X class flare. The X1.4 class flare occurred at 15:37 UT on July 12, which was accompanied by the CME we study in this work. This AR also produced another full halo CME on 2012 July 19, one partial halo CME on 2012 July 17 and the 2012 July 23 extreme space weather event [Ngwira et al., 2013; Russell et al., 2013; Baker et al., 2013].

The full halo CME of interest was first seen in the LASCO C2 field of view on 2012 July l2 at 16:48 UT. Figure 1 shows the running-difference coronagraph images at 16:54 UT and 17:24 on July 12, 2012 from STEREO-A (STA) COR2, STEREO-B (STB) COR2 and SOHO/LASCO C2. The positions of STA and STB in Heliocentric Earth Ecliptic (HEE) coordinates on 2012 July 12 are illustrated in Figure 2.

The projected linear speed, according to the CDAW LASCO CME catalog, is 885 km/s. The X1.4 flare peaked at 16:49 UT with the location at the heliographic coordinate S13W15. Using STEREO observations and the Graduated Cylindrical Shell (GCS) model to reconstruct and measure the 3D CME [Thernisien et al., 2006, 2009], we determined that the propagation direction of the CME is S09W01, which was almost pointing to the Earth. The CME reached 5 Rs from the Sun center at 16:55 UT at a true speed of 1494 km/s [Hess and Zhang, 2014] as determined by fitting the height measurements from the GCS model.

About two days later, this CME arrived on Earth and a strong shock was recorded by the Wind spacecraft on July 14 at 17:00 UT. The interplanetary disturbance caused by this event, the CME-driven shock, took about 48 hours and 12 minutes to reach the Earth.

**2.2 3D MHD Model and Simulation Method**

In this section, the 3D MHD simulation of the background solar wind for Carrington Rotation (CR) 2125 is presented. The computational domain here covers 1 $R_s$ ≤ $r$ ≤ 220 $R_s$; -89° ≤ $\theta$ ≤ 89° and 0° ≤ $\varphi$ ≤ 360°, where $r$ is the radial distance from solar center in units of solar radius $R_s$, and $\theta$ and $\varphi$ are the elevation and azimuthal angles, respectively. The grid mesh is chosen to be 464($r$) × 89($\theta$) × 180 ($\varphi$). The grid size is uniform in azimuth, with $\Delta\varphi$=2°. The radial grid ($r_i$) and meridional grid ($\theta_j$) are not uniform. In order to obtain a precise computational resolution, we choose for the radial grid: $r(1)$=1.0 $R_s$, $\Delta r(1) = s \times r(1)$, $r(i)=r(i-1)+\Delta r(i-1)$, $\Delta r(i)=s \times r(i-1)$, where $s=\pi/225$ ($\pi$=3.1415926) between 1 $R_s$ and 23 $R_s$, and $s=\pi/315$ between 23 $R_s$ and 220 $R_s$. The spatial resolution in the radial direction gradually varies from ~0.01 Rs at the inner boundary of 1 Rs to ~2.0 Rs near 1AU. For the meridional grid we choose $\Delta\theta$ (0°)=1.0°, $\Delta\theta$ (-89°)= $\Delta\theta$ (89°)=3.0°, with a constant increase in $\Delta\theta$ from $\theta$=0° to $\theta$=±89°.

The numerical scheme we used is a 3D corona interplanetary total variation diminishing (COIN-TVD) scheme in a Sun-centered spherical coordinate system ($r$, $\theta$, $\varphi$) [Feng et al., 2003, 2005; Shen et al., 2007, 2009]. The time-dependent 3D ideal MHD equations used in this study include solar rotation [e.g., Shen et al., 2007] and heating source term [Feng et al., 2010; Zhou et al., 2012], where the pressure equation and the volumetric heating function $S_E$ are given by:

$$\frac{\partial p}{\partial t} + \frac{1}{r^2}\frac{\partial r^2(pv_r)}{\partial r} + \frac{1}{r\sin\theta}\frac{\partial \sin\theta(pv_\theta)}{\partial \theta} + \frac{1}{r\sin\theta}\frac{\partial(pv_\varphi)}{\partial \varphi}$$
$$= -(\gamma-1)p\nabla \bullet \vec{v} + (\gamma-1)S_E$$

where $S_E$ is the heating source term, with the form of: $S_E = Q\exp[-r/L_Q]$, which is defined by following the work of [Nakamizo et al., 2009; Feng et al., 2010, 2011; Zhou et al., 2012]. $Q$ and $L_Q$ are the intensity and decay length of heating, respectively. The heating intensity is defined as $Q=Q_0/f_s$. In this research, $L_Q$ and the constant value of $Q_0$ are set to be 0.9 Rs and $1.0\times10^{-6} \mathrm{Jm^{-3}s^{-1}}$, respectively. The expansion factor $f_s$ is defined as:

$$f_S = (\frac{R_s}{r})^2 \frac{B_{R_s}}{B_r},$$ where $R_s$ and $r$ are 1 solar radius and the distance from the solar center, and $B_{Rs}$ and $B_r$ are magnetic field strength at the solar surface and at $r$. In this simulation, the expansion factor is time-invariant and same as the value we used during calculating the background solar wind. The involvement of expansion factor in the heating source term is encouraged by the fact that the solar wind speed is inversely correlated with the expansion rate of the magnetic flux tube in the corona [Levine et al., 1977]. Here, we follow the work by Rempel et al. [2009] to calculate the diffusion flux $f_{i+\frac{1}{2}}$ by using the extrapolated values at the cell interface $u_l = u_i + 0.5\ \Delta u_i$ and $u_r = u_{i+1} - 0.5\ \Delta u_{i+1}$:

$$f_{i+\frac{1}{2}} = \frac{1}{2} c_{i+\frac{1}{2}} (u_r - u_l),$$ where $\Delta u_i$ is limited by the slope limiter minmod to make the numerical scheme TVD, as shown following [Feng et al., 2003]:

$$\Delta u_i = \min \mathrm{mod}(\delta u_{i-1/2}, \delta u_{i+1/2}),\ \delta u_{i+1/2} = u_{i+1} - u_i,\ \delta u_{i-1/2} = u_i - u_{i-1},$$

Here, $\mathrm{minmod}(x, y) = \mathrm{sgn}(x) \mathrm{Max}(0, \min[|x|, y\mathrm{sgn}(x)])$, where $\mathrm{sgn}(x) = \begin{cases} 1, & x > 0; \\ 0, & x = 0; \\ -1, & x < 0. \end{cases}$

The characteristic velocity $c$ is defined as $c = 0.1\ c_{\mathrm{sound}} + v + v_{\mathrm{alf}}$, which significantly reduces the diffusivity in low Mach number flows. This scheme has been applied to all MHD variables and account for the effects of mass diffusion in the momentum and pressure fluxes. The $\nabla \bullet \vec{B}$ error produced by the diffusion scheme is controlled by iterating:

$$\vec{B}^{n+1} - \vec{B}^n = \mu(\Delta x)^2 grad(div\vec{B}^n),$$

where *n* is the number of iteration and $(\Delta x)^2 = \dfrac{3}{\dfrac{1}{(\Delta r)^2} + \dfrac{1}{(r\Delta\theta)^2} + \dfrac{1}{(r\sin\theta\Delta\varphi)^2}}$ is in the spherical coordinate system. The value $\mu$ is set as 0.35 to satisfy $\max(\dfrac{\Delta x |\nabla \bullet \vec{B}|}{|\vec{B}|}) < 10^{-3}$ in less than 50 iterations (*n*<50). This artificial diffusivity can lead to a scheme that is fully shock-capturing, at least second-order accuracy in smooth regions (higher order is possible depending on the slope limiter used) [Rempel et al., 2009; van der Holst and Keppen, 2007; Feng et al., 2011 ].

At the inner boundary (1 Rs), the method of projected characteristics [Wu and Wang, 1987; Hayashi, 2005; Wu et al., 2006] is employed. At the outer boundary of *r*=220 Rs, and the boundaries at -89° and 89°, we employ a linear extrapolation. The detailed description of the asynchronous and parallel time-marching method for the 3D MHD simulation are discussed in detail in Shen et al. [2007, 2009, 2011a, 2011b].

We first establish a steady state background solar wind. The potential field, extrapolated from the observed line-of-sight magnetic field of CR 2125 on the photosphere from the Wilcox Solar Observatory (WSO), and Parker's solar wind solution are used as the initial magnetic field and velocity. The initial density is deduced from the momentum conservation law, and the initial temperature is given by assuming an adiabatic process. With these initial conditions, our MHD code may quickly reach a self-consistent steady state of solar wind.

Figure 3 shows the steady-state distribution of radial-component of magnetic field and velocity at 5 Rs. The location of $B_r = 0$ is indicated by the dashed lines. Figure 3 indicates that the corona current sheet becomes nearly vertical to the ecliptic plane, which is rather typical at the solar maximum. From the bottom panel of Figure 3, it could also be found that the distribution of the low-speed region is basically consistent with the corona current sheet region.

The CME is modeled as a magnetic blob with its center sitting at $r = 5$ Rs, just as we did in the previous work [Chané et al., 2005; Shen et al., 2011a; Shen et al., 2013]. To reproduce the evolution of the 2012 July 12 event, the initial propagation direction and velocity are chosen to be the same as those determined from observations. From the observations, the direction of the CME is S09W01, and the propagation speed at 5 Rs is 1494 km/s. Thus, the average speed of the plasma blob ($v_{ave}$) is set to be 1494 km/s and the maximum velocity inside the plasma blob should be $\sim 3 v_{ave}$ [Chané et al., 2005; Shen et al., 2011a].

The density, radial velocity and temperature profile of the initial perturbation are defined as follows:

$$\begin{cases} \rho_{CME}(r,\theta,\varphi) = \frac{\rho_{\max}}{2}(1-\cos(\pi \frac{a_{cme}-a(r,\theta,\varphi)}{a_{cme}})) \\ V_{CME}(r,\theta,\varphi) = \frac{v_{\max}}{2}(1-\cos(\pi \frac{a_{cme}-a(r,\theta,\varphi)}{a_{cme}})) \\ T_{CME}(r,\theta,\varphi) = \frac{T_{\max}}{2}(1-\cos(\pi \frac{a_{cme}-a(r,\theta,\varphi)}{a_{cme}})) \end{cases}$$

where $a_{cme}$ is the radius of the initial plasma blob, $a(r, \theta, \varphi)$ denotes the distance from the center of the initial plasma blob, $\rho_{\max}$, $v_{\max}$ and $T_{\max}$ are the maximum density, radial velocity and temperature in the plasma bubble added on top of the background solar wind, respectively.

The initial magnetic field of the perturbation in $r$ and $\theta$ direction can be defined as [Shen et al., 2011a, 2011b]:

$$\begin{cases} B_{r_{CME}}(r,\theta,\varphi) = -\frac{1}{r^2 \sin\theta}\frac{\partial \psi(r,\theta,\varphi)}{\partial \theta} \\ B_{\theta_{CME}}(r,\theta,\varphi) = \frac{1}{r \sin\theta}\frac{\partial \psi(r,\theta,\varphi)}{\partial r} \end{cases}$$

where

$$\psi(r,\theta,\varphi) = \psi_0 (a(r,\theta,\varphi) - \frac{a_{CME}}{2\pi}\sin(\frac{2\pi a(r,\theta,\varphi)}{a_{CME}}))$$

is the magnetic flux function. $\Psi_0$ is constant, and different sign of $\Psi_0$ denotes different polarity of the magnetized plasma blob [Chané et al., 2006]. The two panels of Figure 4 give the 3D views of the CME initialization, showing the iso-surfaces of radial velocity ($v_r$) and the magnetic field lines by using different $\Psi_0$ of -4.0 (a) and 4.0 (b). It could be found that the polarity of the initial CMEs in panel (a) and (b) is opposite.

Table 1 lists the initial parameters of the CME from observations. The choice of other parameters is given to match the transit time of the shock, the total magnetic field, and other Wind data at the shock as the best fit as possible. Therefore, all the initial parameters of the CME are constrained by observations. The input of the mass and the momentum of the CME are $2.45\times10^{12}$kg and $3.67\times10^{15}$kgkms$^{-1}$, respectively. The relative pressure, which is defined as ($P_{CME}$-$P_{bg}$)/ $P_{bg}$, is about 0.32, where $P_{CME}$ and $P_{bg}$ are the pressures of CME and local background solar wind, respectively.

Table 1. Initial parameters of the CME

|  | D | $v_{ave}$ km/s | $n_{ave}$ $\times 10^7$cm$^{-3}$ | $T_{ave}$ $\times 10^6$ K | $\Psi_0$ | $a_{CME}$ Rs | $B_{ave}$ $\times 10^5$ nT | Mass $\times 10^{12}$kg | Momentum $\times 10^{15}$kgkms$^{-1}$ |
|---|---|---|---|---|---|---|---|---|---|
| CME | S09W01 | 1494 | 2.0 | 6.0 | -4.0 | 0.6 | 6.0 | 2.45 | 3.67 |

## 3. Kinematic evolution of the CME

### 3.1 Comparisons with Coronagraph images

Detailed studies of synthetic line-of-sight images from modeling have been performed by a number of groups in the past [Chen and Krall, 2003; Manchester et al., 2004; Lugaz et al., 2005, 2009; Odstrcil et al., 2005; Riley et al., 2008]. The direct comparison of such synthetic observations with real observations has only been done in recent years for a few selected events during or close to the solar maximum [Lugaz et al., 2007, 2009; Manchester et al., 2008; Sun et al., 2008].

Synthetic coronagraph images of CMEs seem to be a simple and relevant way to display 2D (two-dimensional) representations of simulated CMEs [Lugaz et al., 2009]. Line-of-sight images are the best way to study the density structure of CMEs. Producing synthetic white-light images and comparing those with 3D data sets will provide information on how the density structure of a CME obtained from real coronagraphs is related to the 3D structure of the CME [Lugaz et al., 2007].

To turn the 3D simulation into an image comparable to remote sensing images, the approximate position of the desired view (SOHO, STEREO-A and B) must be determined in a Heliocentric Coordinate system. The positions of STA and STB in HEE coordinates at 2012 July 12 are plotted in Figure 2. At that time, STA is separated away from the Earth by about 120º at a radial distance of 0.96 AU from the Sun and STB by about 115° at a radial distance of 1.01 AU from the Sun. The total field of view must also be specified, in terms of an angular extent. For instance, the field of view on the STEREO COR-2 instrument is approximately 2-15 Rs, or in terms of angular field of view, approximately 0.5-4° from the solar center. For each pixel in the computation domain, its angular position relative to the observer is calculated. If the pixel is within the angular limits of the image cone, it is projected onto the plane of sky and the image coordinates of the pixel are calculated. The value of the relative density of the computational pixel is then added into the image pixel. This calculation is carried out over each pixel in the grid and the sums are compiled for the each image pixel so that a complete image can be assembled.

Figure 5 (a) shows remote sensing observations from STEREO COR2B, COR2A and SOHO/LASCO C2 at 17:24 UT on July 12. Figure 5(b) to 5(e) show the synthetic images from simulation at 17:24 UT (b) and 21:54 UT (c) on July 12, at 08:54 UT (d) on July 13 and 08:54 UT (e) on July 14, respectively. The synthetic images are produced based on the relative density data from the 3D simulation output. Figure 6 shows the comparison of real and synthetic SECCHI/HI-1-B (left) and SECCHI/HI-1-A (right) images at 21:29 UT on July 12. There is a fair agreement of the overall shape and the propagation direction of the CME between observations and simulations. However, the CME leading edge in the synthetic images moves slightly more ahead than that in the real coronagraph images. This might be due to the adopted CME blob, since the unrealistic initiation mechanism is challenging unclear. The CME model doesn't include the expansion speed at its initialization. Therefore, at the early stage, the simulated CME propagate a bit faster than the observations in $r$-direction, while in the direction perpendicular to *r*-direction, the expansion of the simulated CME is quite smaller than that of the observations.

**3.2 Comparisons with Time-Elongation Maps (J-maps) and Kinetic Evolution of the Shock**

The comparison between synthetic numerical results and real white-light observations in Section 3.1 is mainly for the CME morphology at very early stages. The following comparison is for the CME kinematics from near the Sun to 1AU. One of the methods to track the CME in interplanetary space is to produce time-elongation maps (J-maps) [e.g. Sheeley et al., 1997; Sheeley et al., 2008; Rouillard et al., 2008; Davies et al., 2009]. J-maps allow for the tracking of CMEs to large elongation angles and enable the study of their evolution without concerning the direction of propagation.

Here, we study the J-maps along the Sun-Earth line. Considering the Thomson Scatter [Jackson, 1997], we translate the simulated density distribution in the ecliptic plane to the brightness distribution. Then, a slice is obtained by showing the total brightness along the elongation angles from 1° to 80°. To get the synthetic J-map, we take a total brightness of the slice every 30 minutes and plot the running-difference results. For the real J-map, a slice is taken for every observational image and the running difference is plotted.

Figure 7(a) and 7(b) give the synthetic J-maps corresponding to the position of STA and STB, respectively. Figure 7(c) and 7(d) show the real J-maps constructed from the imaging data from COR2, HI1 and HI2 imagers onboard STA and STB by placing a slice along the ecliptic plane. Every stripe in a J-map indicates a featured element moving away from near the Sun to 1AU. In order to compare the stripes of the CME feature between the synthetic J-maps and real J-maps, we track and locate the CME's leading edge in the synthetic J-maps by red diamonds (Figure 7(a) and 7(b)) and in the real J-maps by blue diamonds (Figure 7(c) and 7(d)). The real J-map at the STA is much more complicated and harder to recognize at large elongation, so we only mark the CME's leading edge in the region with elongation less than 23º, as shown in Figure 7(c). Then, we make the quantitative comparison between synthetic numerical results and real white-light observations from the J-maps. Figure 8 plots the time-elongation profiles from the observations (blue diamond) and the synthetic images derived from simulation (red diamond) corresponding to the position of STA and STB. This comparison shows that the simulation offers a satisfactory reproduction of the observations, except at very early stage, e.g., before 22:00UT on July 12.

The time of introducing CME into the computational domain is set to be zero. We locate the CME by simply setting a threshold of 0.5 in the map of relative density $(\rho - \rho_0)/\rho_0$, where $\rho$ is the total density, and $\rho_0$ is the density of the background solar wind. Figures 9(a) to 9(d) show the 3-D view of the relative density and magnetic field distribution at $t$ = 0.5, 10, 20, and 30 hours. The color code in the panels represents the two level isosurfaces of the relative density. And the outer isosurface with the relative density of 0.5 mainly denotes the shock surface of the CME. The magnetic field topology in Figure 9 is represented by white magnetic field lines. A CME leading edge with a high density is clearly visible in front of the flux rope. At the early time of 0.5 hours, because the initial radial velocity of CME is much larger than the background solar wind speed, the shape of the CME looks like an "olive". As the CME propagates into the heliosphere, it expands obviously in the direction perpendicular to the propagation direction and it is radially compressed.

Next, we focus on the quantitative comparison of the time-heliosdistance and the time-speed on the shock front between the 3D numerical results and the observations. Figure 10 shows the evolution of the relative density ($(\rho-\rho_0)/\rho_0$) - distance profile along the Sun-Earth line at six consecutive times, in which we could recognize that the leading edge of the CME is located near the position of CME density peak. At the very early time of $t$=1 hour, the relative density profile has an obvious sharp jump from ~11 Rs to ~16 Rs. As time goes, the width of the relative density jump along the Sun-Earth line increases apparently. At 10 hours, the width of the jump is near 18 Rs; at 20 hours, the width is ~23 Rs and at 40 hours, the width increases to ~33 Rs. Thus, as the CME propagates into the heliosphere, it also expands in radial direction. We also notice that after the CME passed, the relative density behind the CME drops to negative (but the actual density remains positive), which is probably because the CME removes some of the background's mass when it propagates into the background solar wind.

From Figure 10, we find that the density changes very sharply at the CME's front edge, and we suppose that the shock front is located at the position with the maximum of the density gradient along the propagation direction in front of the CME. The blue dashed lines in Figure 11 show the time-height and the time-speed distribution of the shock front from the simulation. The green dash-dotted lines in Figure 11 give the height and the velocity of the shock front distribution deduced from the observations; to obtain the smooth distribution, we fit the observational data points in the shock drag model [Vršnak et al., 2013; Hess and Zhang, 2014]. And the red diamonds in Figure 11 mark the height and the velocity of shock front distribution from the HM triangulation [Lugaz et al., 2010] based on the observations. From the comparison, we find that at the initial time within 6 hours, the shock front deduced from simulation moves faster than that observed; then, the shock speed from simulation drops quickly and become similar to the observed shock speed. The fast shock may be caused by the large average speed of the plasma blob which was set to be 1494km/s to fit the initial CME speed from the imaging data.

### 3.3 Comparisons with the Wind data at 1AU

Figure 12 depicts the plots of total, $x$-component, $y$-component, $z$-component magnetic field ($B_x$, $B_y$, $B_z$) at the GSE coordinate system, velocity, number density and temperature at 1AU, respectively, from the top to the bottom panels. Each panel describes the comparison of the simulated plasma parameters with the Wind observed parameters. Figure 12 demonstrates that our data-constrained simulation can reproduce well some of the *in-situ* measurements: the transit time of the shock about 48 hours is approximately reproduced, the velocity, the total magnetic field, the temperature and the density peak value are very close to the realistic values during the peak period. The leading shock, characterized by a sharp jump in the total magnetic field, $B_x$, $B_y$, $B_z$, velocity and temperature curves, arrives almost at the same time between the simulation and the in-situ measurements.

Nevertheless, some quantitative disagreement is expected when compare simulation results with real observations. The blue vertical solid lines indicate the arrival time at 1AU of the shock and the blue vertical dashed lines denote the arrival time of the magnetic cloud, which was deduced from the *in-situ* observations [Möstl et al., 2014; Hess and Zhang, 2014]. The shock arrival at 1AU at 1724UT on July 14, immediately followed by a sheath region. At about 06:00UT on July 15, the sheath region ended and a magnetic cloud began [Möstl et al., 2014]. The green vertical solid lines indicate the shock arrival time from simulation, which is only ~1 hour earlier than the observations.

The magnetic cloud in the simulation reaches 1AU at about 22:30UT on July 14, marked by the green dashed lines, which is ~7.5 hours earlier than the observed one. One reason about the disagreement is that the simulated flux rope size is not as large as the observed one. This is mainly caused by the limitation of the model, which assumes a magnetized plasma blob as the flux rope.

## 4. Summary and Discussion

We have investigated the evolution of the June 12-16, 2012 CME in a realistic ambient solar wind by using the 3D data-constrained COIN-TVD MHD simulation. We first established a steady state background solar wind from solar surface to the Earth's orbit (215 $R_s$) and beyond by using the observed line-of-sight magnetic field of CR 2125 on the photosphere. Our numerical results of the background solar wind show that the current sheet becomes nearly vertical to the ecliptic plane, which demonstrates the typical characteristics at solar maximum.

We simulated the CME by means of a high-density, -velocity and -temperature magnetized plasma blob, which is superimposed on the background steady state solar wind. To reproduce the 2012 July 12 event, we chose the initial propagation direction and average velocity to be the same as those derived from observations. The choice of other parameters is given to match the transit time of the shock, the total magnetic field, and other in-situ data at the shock as the best fit as possible.

From the comparisons with remote sensing observations, the J-map versus observations, as shown from Figure 5 to Figure 10, we find that (1) we are able to reproduce successfully the observations in STA and STB field-of-view, for both the CME morphology and the CME kinematics; (2) our results for the shock front propagation are mainly consistent with the results from the shock drag model, except at the very early time.

When the CME evolves to ICME reaching 1 AU, its physical parameters (Figure 12) resemble the observations of the ICME recorded by the Wind spacecraft. Comparing our simulation results with the *in-situ* data, we find that the transit time of the shock is approximately reproduced, the velocity, the total magnetic field, the temperature and the density peak value are very close to the realistic values during the peak period. While there still exist some quantitative disagreement when compare simulation results with real observations, especially during the interval of a magnetic cloud. The possible reasons might be the uncertainty of the initial realistic solar wind speed and the IMF background conditions and uncertainty of the appropriate solar observations used to initiate the CME.

Besides, from studying the shock front speed-time distribution as shown in Figure 11, we find that at the initial time within 6 hours, the shock front appears to decelerate very quickly from ~2300 km/s to ~1100 km/s; and at 6 hours, the height of the shock front reaches up to ~56 Rs. Then from 56 Rs to 1 AU, the shock front decelerates slowly from ~ 1100 km/s to ~570 km/s. It has been found that the major force which caused the deceleration of a shock or CME is the aerodynamic drag force,

$F_D = -\rho_e A C_D (V_i - V_e) |V_i - V_e| / \tau$, where $\tau$ and $A$ are the volume and the cross-sectional area of the CME, $C_D$ is the drag coefficient, $V_i$ is the CME speed, and $\rho_e$ and $V_e$ are the density and speed of the background solar wind, respectively [Cargill, 2004; Cargill et al., 1994; Schmidt and Cargill, 2000; Temmer et al., 2012, Shen et al., 2012; Vršnak and Gopalswamy, 2002; Owens and Cargill, 2004; Vršnak et al., 2012]. Therefore, the choice of different solar wind speeds would affect the CME's transit time. Actually, Heinemann [2002] has also demonstrated that two major sources of uncertainty in estimates of shock arrival times were the velocity and the density of the ambient medium.

In order to discuss the influence of the background solar wind speed on the shock arrival time (SAT) quantitatively, we have made a test by using the background solar wind with different speed through which the CME propagates. The average background solar wind speed ($V_{sw}$) against the heights in nine different cases are presented from SW1 to SW9 in Figure 13(a). Then, the CME model which was described in section 2.2 is input into the different background solar wind. Figure 13(b) gives the SAT in the nines cases against $V_{sw}$ at 1AU. It could be found that the SAT is almost inversely proportional to the background solar wind. As the ambient solar wind speed at 1 AU increases from 300 km/s to 500 km/s, the transient time decreases from 57 hours to 37 hours. In other words, the slower the solar wind speed, the larger the aerodynamic drag force, the larger the deceleration of the CME, and thus the longer the transit time, which is consistent with the results by Heinemann [2002].

In summary, we have demonstrated that the data-constrained 3D MHD simulation can reproduce the realistic observations to a large extent: not only the arrival time, but also the continuous kinematic process and morphological scenario structures from the Sun to the Earth. This study also reveals certain limitations of the numerical model, such as the less extension of the simulated flux rope. Further refinement of the numerical model is needed, in order to fully simulate the observations.

Moreover, in our present numerical CME model, like many other numerical CME models, there exist two extremely important and still unsolved issues: the uncertainty of the initial realistic solar wind background conditions and the uncertainty of the appropriate solar observations used to "mimicking" solar flare/filament and CME initiation [Dryer 1998; Fry et al., 2001; Odstrcil et al., 2004; Shen et al., 2007, 2011]. To some extent, our objective of using more observational data such as the photospheric magnetic fields by constraining the model is to try to reduce the uncertainty in the initial values of realistic solar wind. But, it is still a challenging problem that how to use the approximate solar observations to initialize the solar flare/filament and CME.

**Acknowledgements**. The data for this work are available at the official websites of STEREO, SOHO and Wind spacecraft. We acknowledge the use of them. STEREO is the third mission in NASA's Solar Terrestrial Probes program, and SOHO is a mission of international cooperation between ESA and NASA. The STEREO/SECCHI data are produced by a consortium of NRL (US), LMSAL (US), NASA/ GSFC (US), RAL (UK), UBHAM (UK),MPS (Germany), CSL (Belgium), IOTA (France), and IAS (France). The Wind data are obtained from the GSFC/SPDF OMNI Web interface at http://omniweb.gsfc.nasa.gov. The Wilcox Solar Observatory (WSO) data used in this study were obtained via the Web site http://wso.stanford.edu/synopticl.html for CR 2125. The WSO is currently supported by NASA. This work is jointly supported by grants from the 973 key projects (2012CB825601, 2011CB811403), the Knowledge Innovation Program of the Chinese Academy of Sciences (KZZD-EW-01-4), the National Natural Science Foundation of China (41231068, 41174150, 41274192, 41131065, 41121003, 41274173 and 41474152), and the Specialized Research Fund for State Key Laboratories. J.Z. and P. H is supported by NSF grants ATM-0748003 and AGS-1156120. We are very grateful to two anonymous reviewers for their constructive and helpful comments.

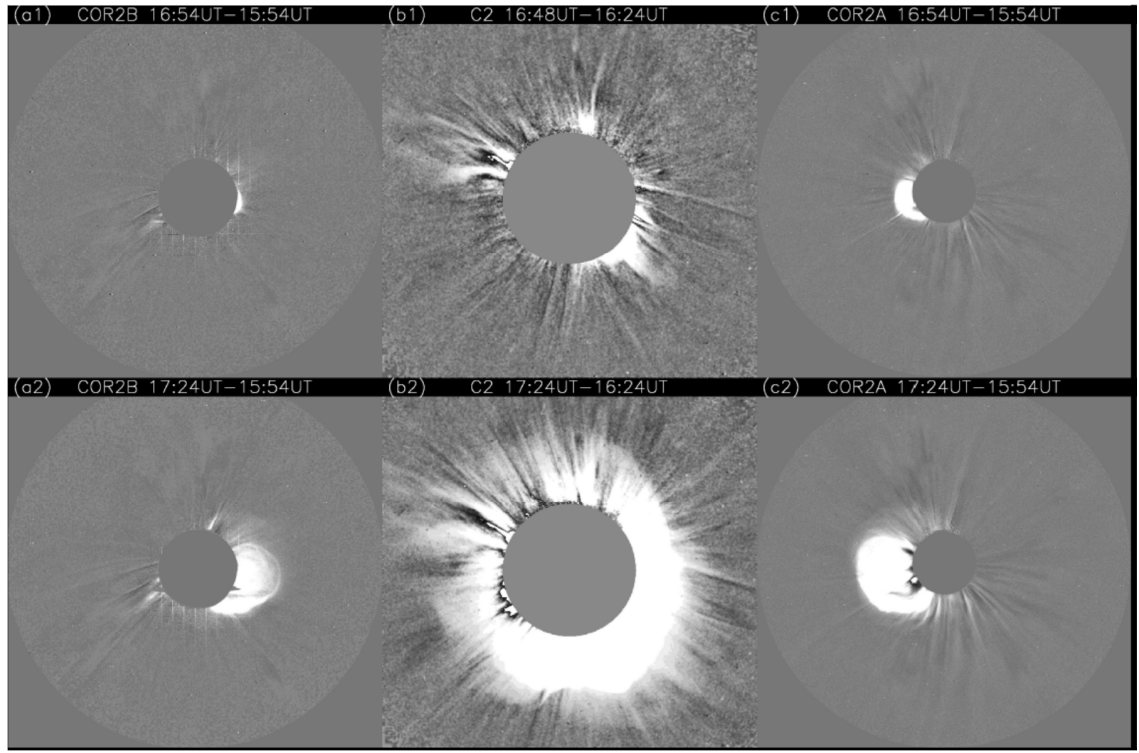

**Figure 1**. Running difference images at 16:54 UT (top) and 17:24 UT (bottom) on July 12, 2012 from STEREO-A COR2 (Left), STEREO-B COR2 (Right) and SOHO/LASCO C2 (Middle), respectively.

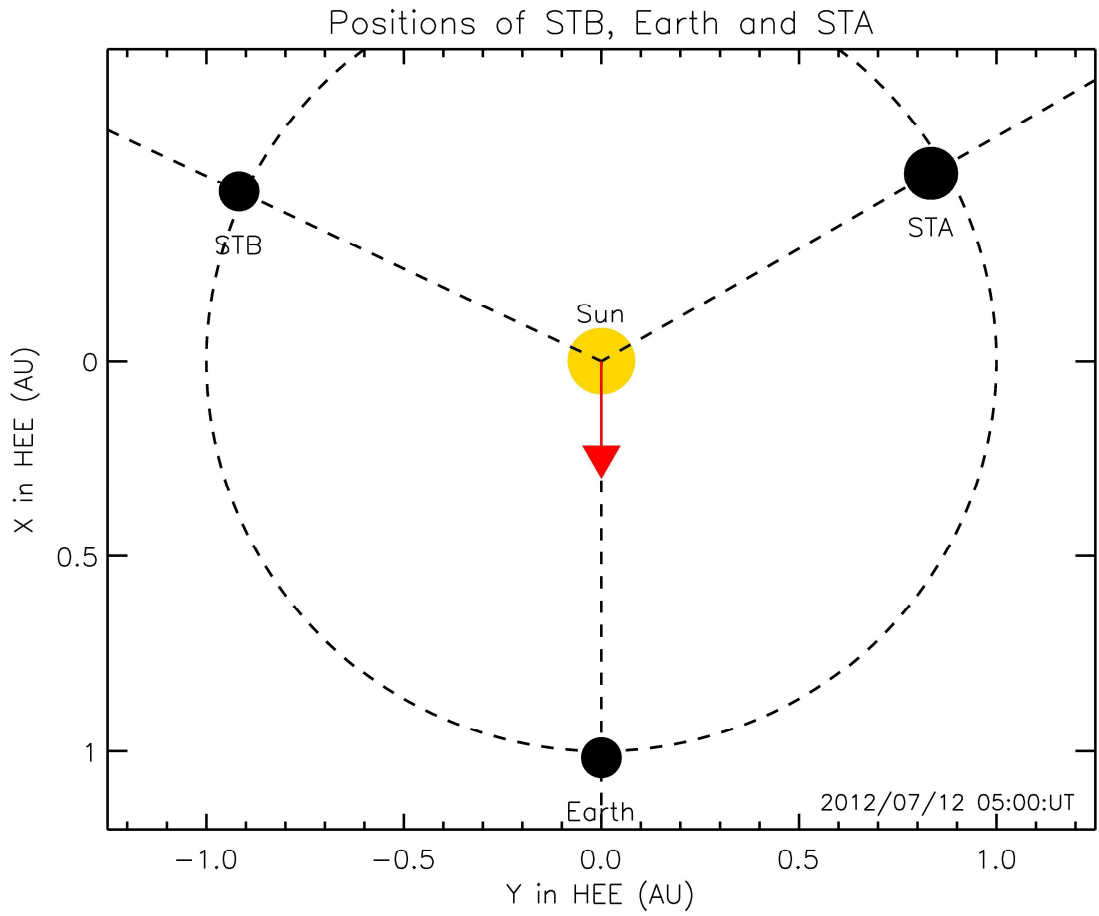

**Figure 2**. The positions of STEREO-A (STA) and STEREO-B (STB) in HEE coordinates on 2012 July 12.

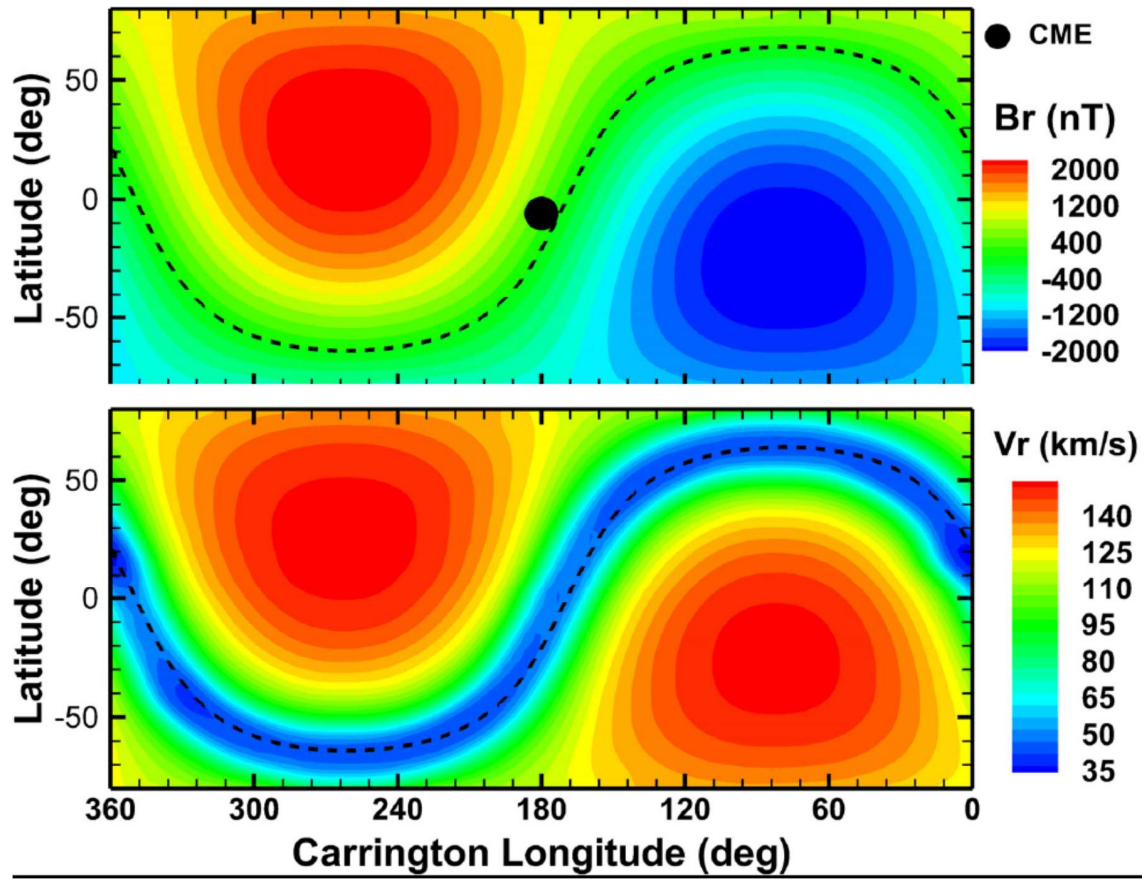

**Figure 3**. The steady-state distribution of radial-component of magnetic field (top) and velocity (bottom) at 5 Rs. The dashed lines in each panels show the location of $B_r = 0$.

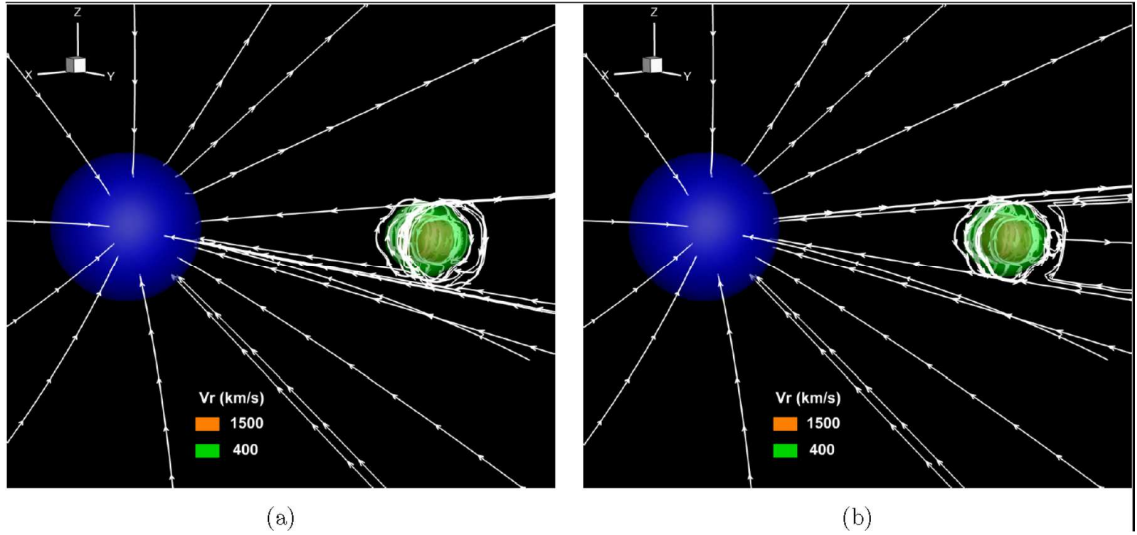

**Figure 4**. 3D views of the CME initialization, including two levels of iso-surface of the radial velocity and the magnetic field lines with (a) $\Psi_0$= -4.0 and (b) $\Psi_0$= 4.0.

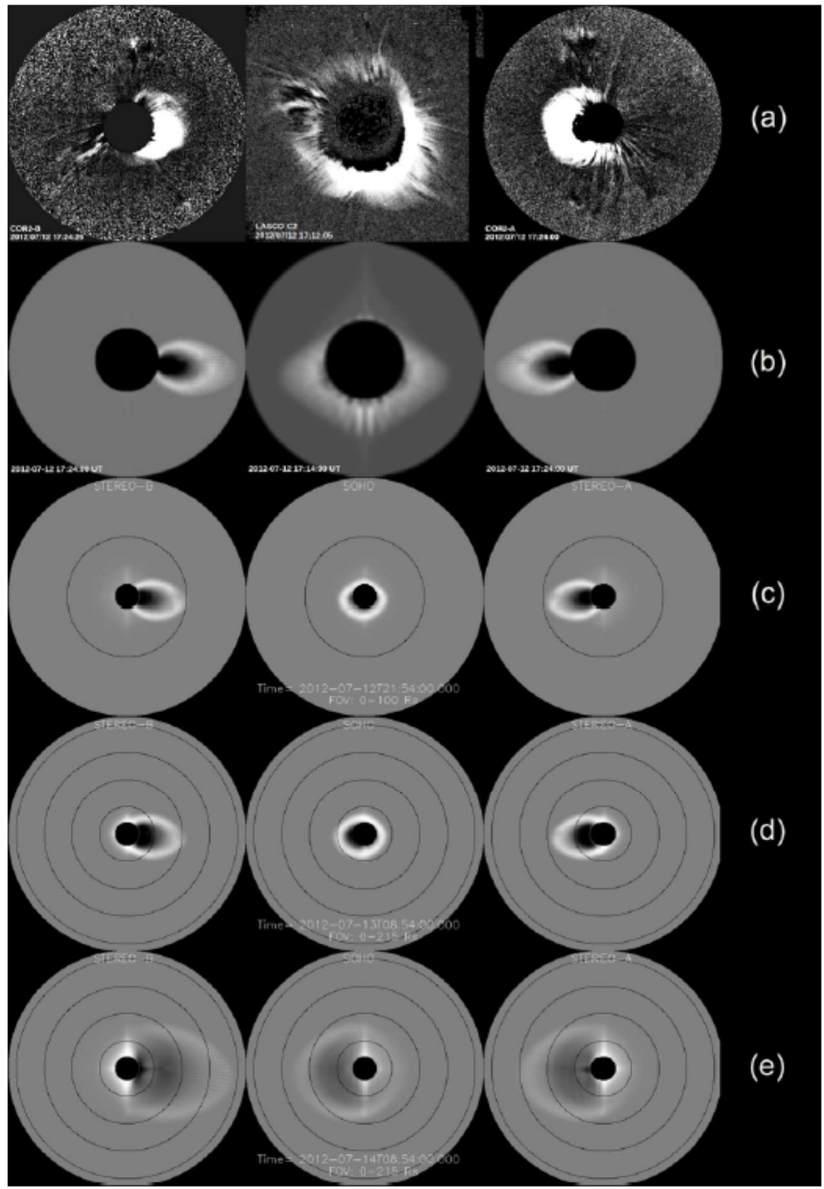

**Figure 5.** Comparison of real (a) and synthetic (b) STEREO/COR2B (left), STEREO/COR2A (right) and LASCO/C2 (middle) images at 17:24 UT on July 12; (c)-(e): Synthetic images at 21:54 UT (c) on July 12, at 08:54 UT (d) on July 13 and at 08:54 UT (e) on July 14.

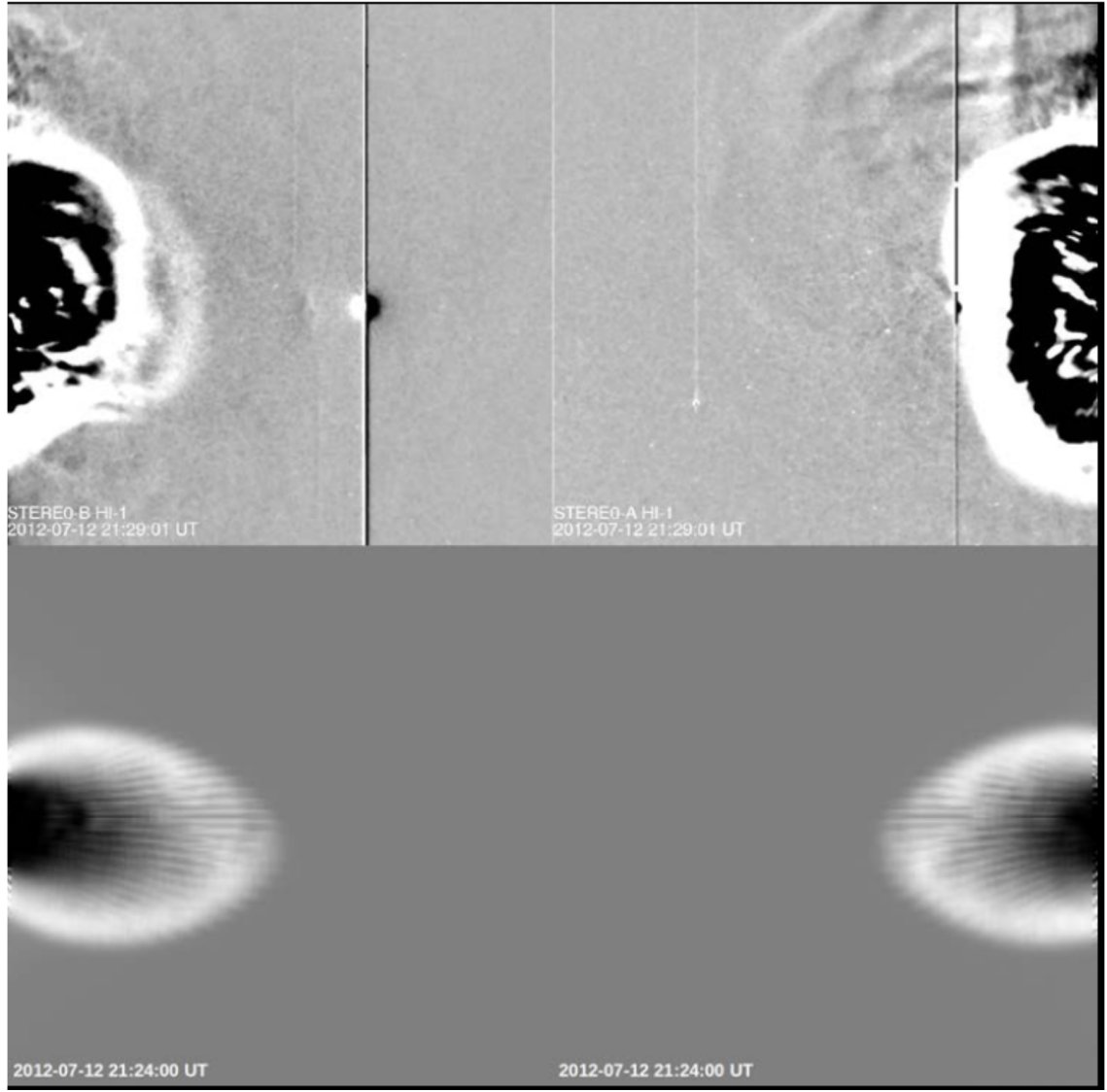

**Figure 6**. Comparison of real (top) and synthetic (bottom) SECCHI/HI-1-B (left) and SECCHI/HI-1-A (right) images at 21:29 UT on July 12.

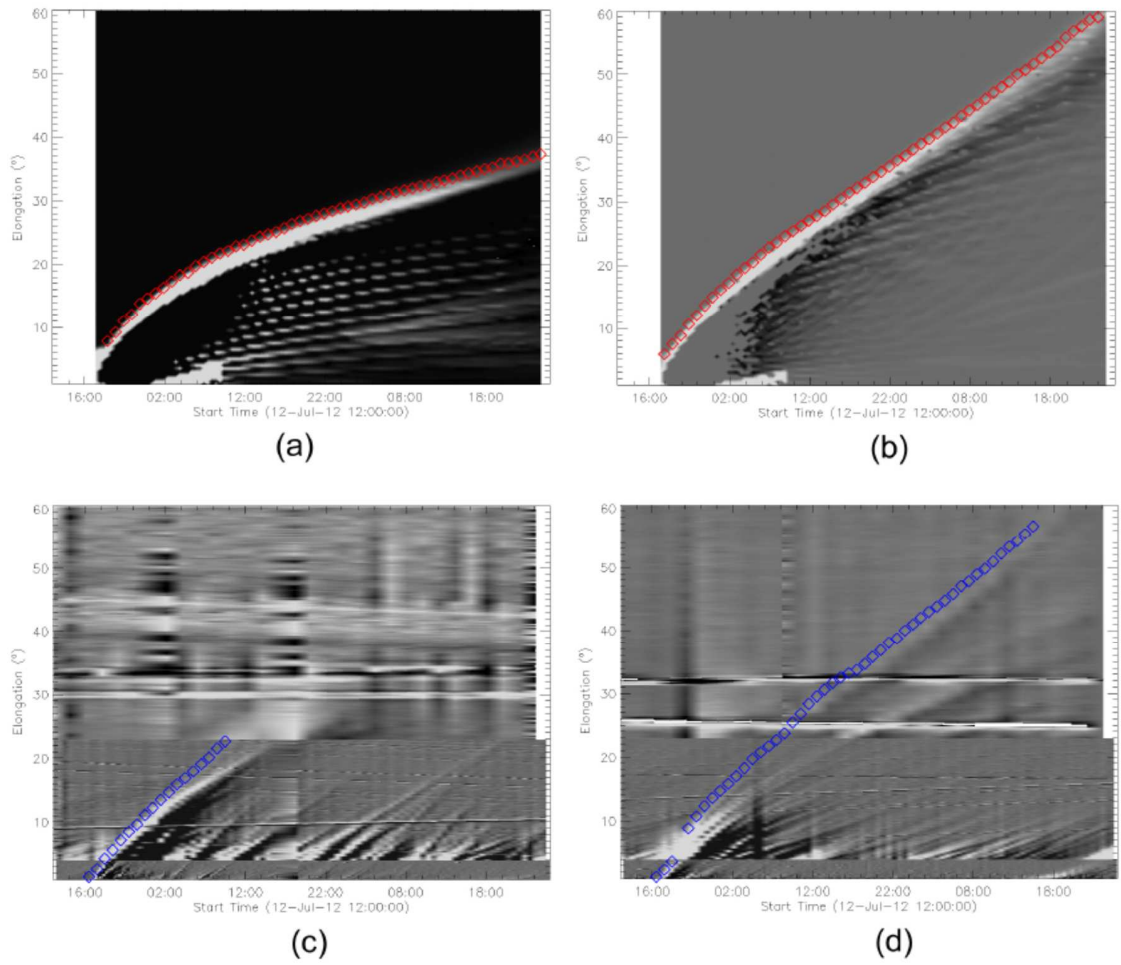

**Figure 7**. (a) and (b): Synthetic J-maps corresponding to the position of STA and STB; (c) and (d): Real J-maps constructed based on the imaging data from COR2, HI1 and HI2 imagers onboard STA and STB.

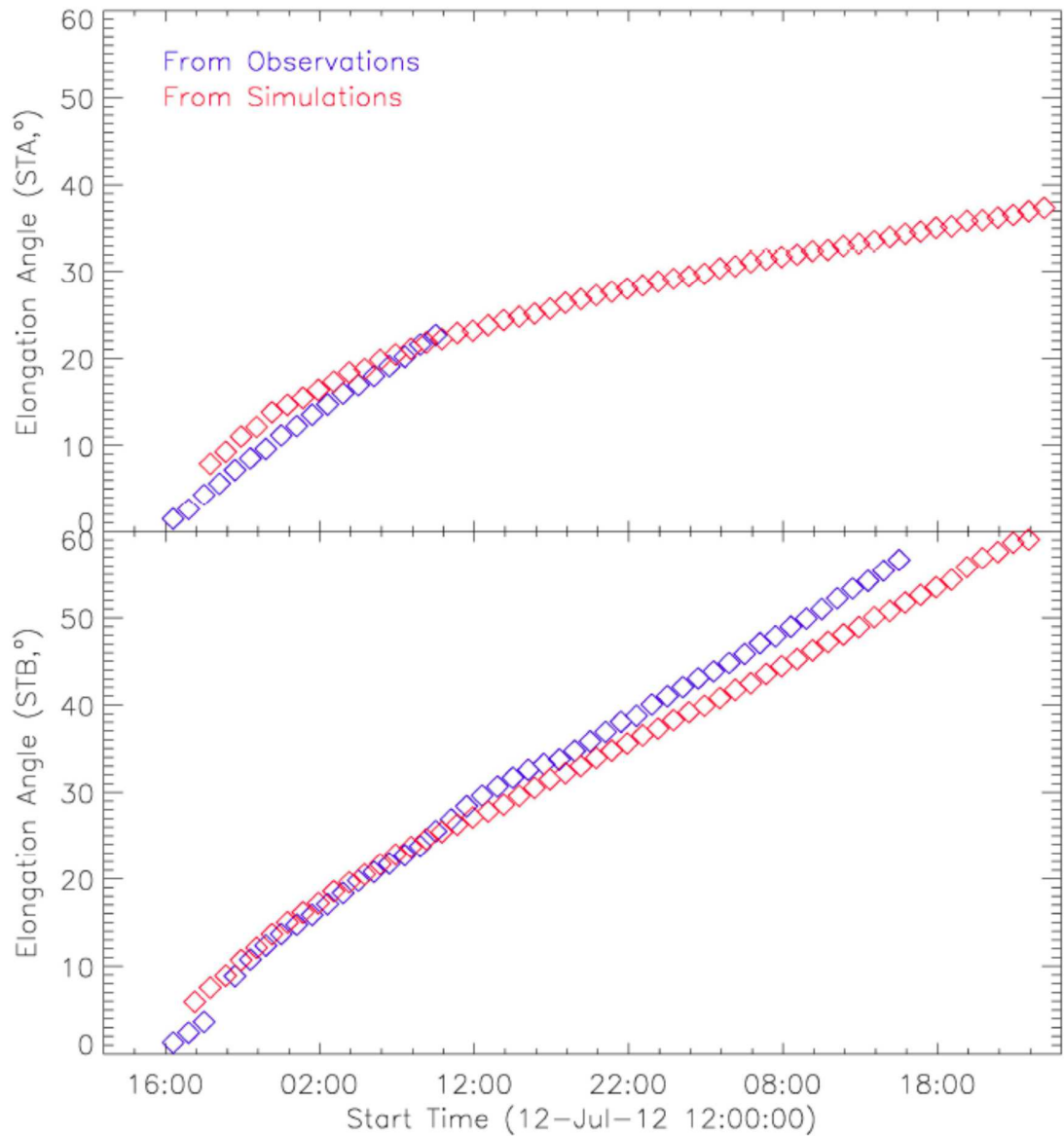

**Figure 8**. Comparison between time-elongation profiles from the observations (blue diamond) and the synthetic images derived from simulation (red diamond) corresponding to the position of STA (top) and STB (bottom).

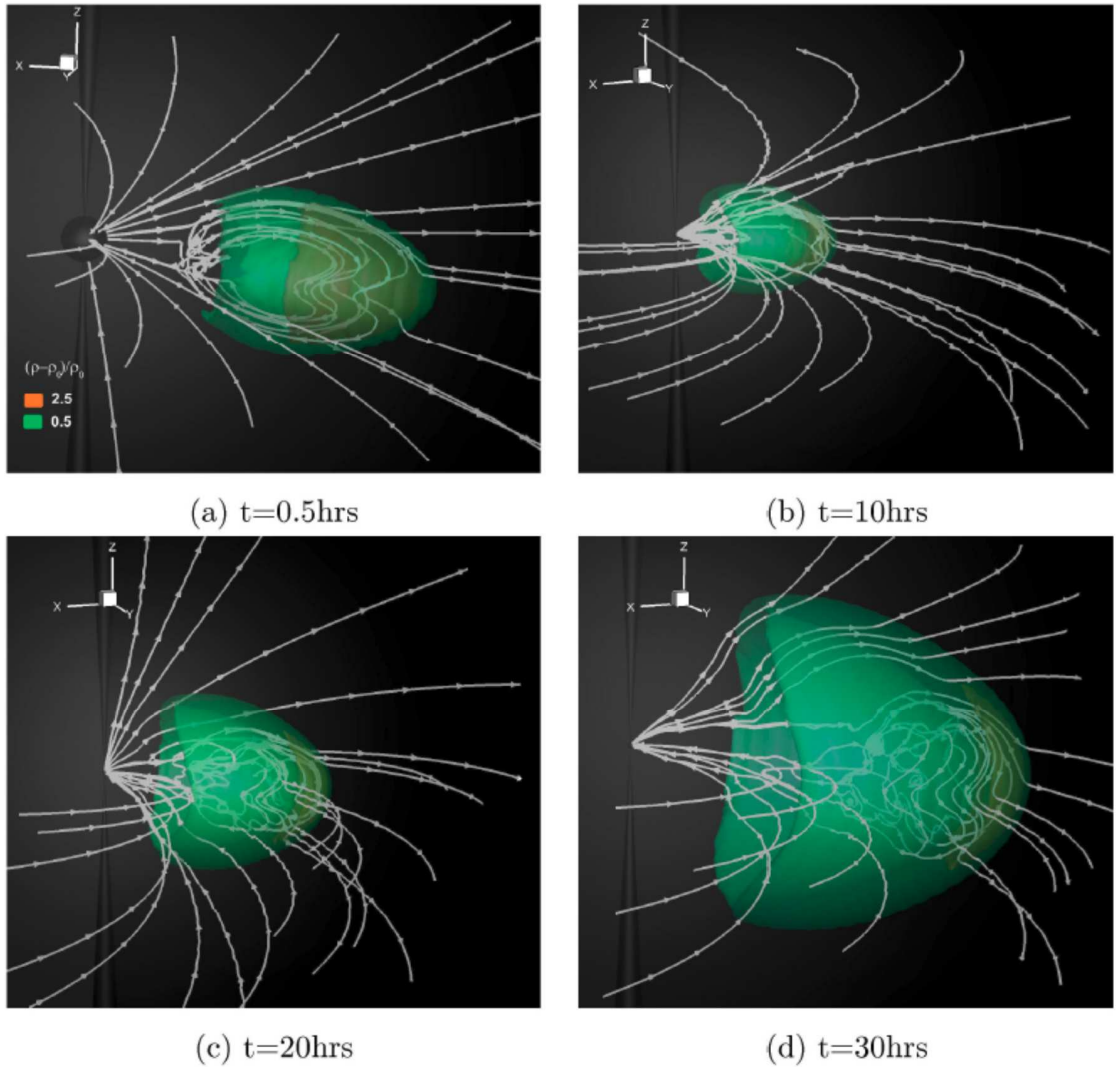

**Figure 9**. 3-D view of the relative density ($(\rho - \rho_0)/\rho_0$) distribution at t = 0.5, 10, 20, and 30 hours. The color code in the panels represents the two levels isosurfaces of relative density. The magnetic field topology is represented by the white magnetic field lines.

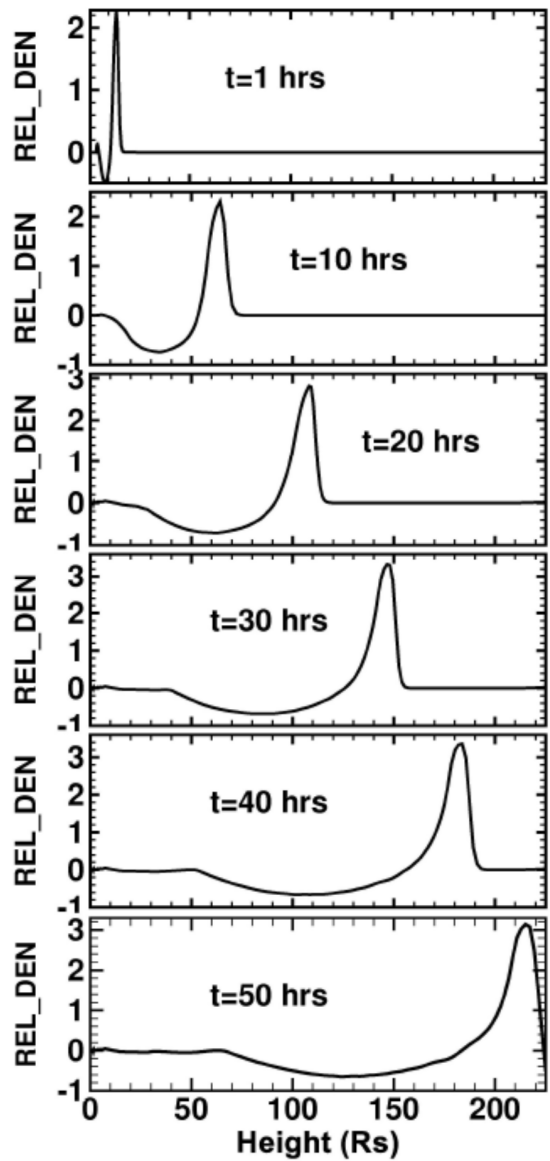

**Figure 10.** Evolution of the relative density ($(\rho - \rho_0)/\rho_0$) vs. heliocentric distance from 0 to 225 Rs along the Sun-Earth line ($\theta=0°$ and $\varphi=180°$) at, (from top to bottom) $t$=1, 10, 20, 30, 40 and 50 hours, respectively.

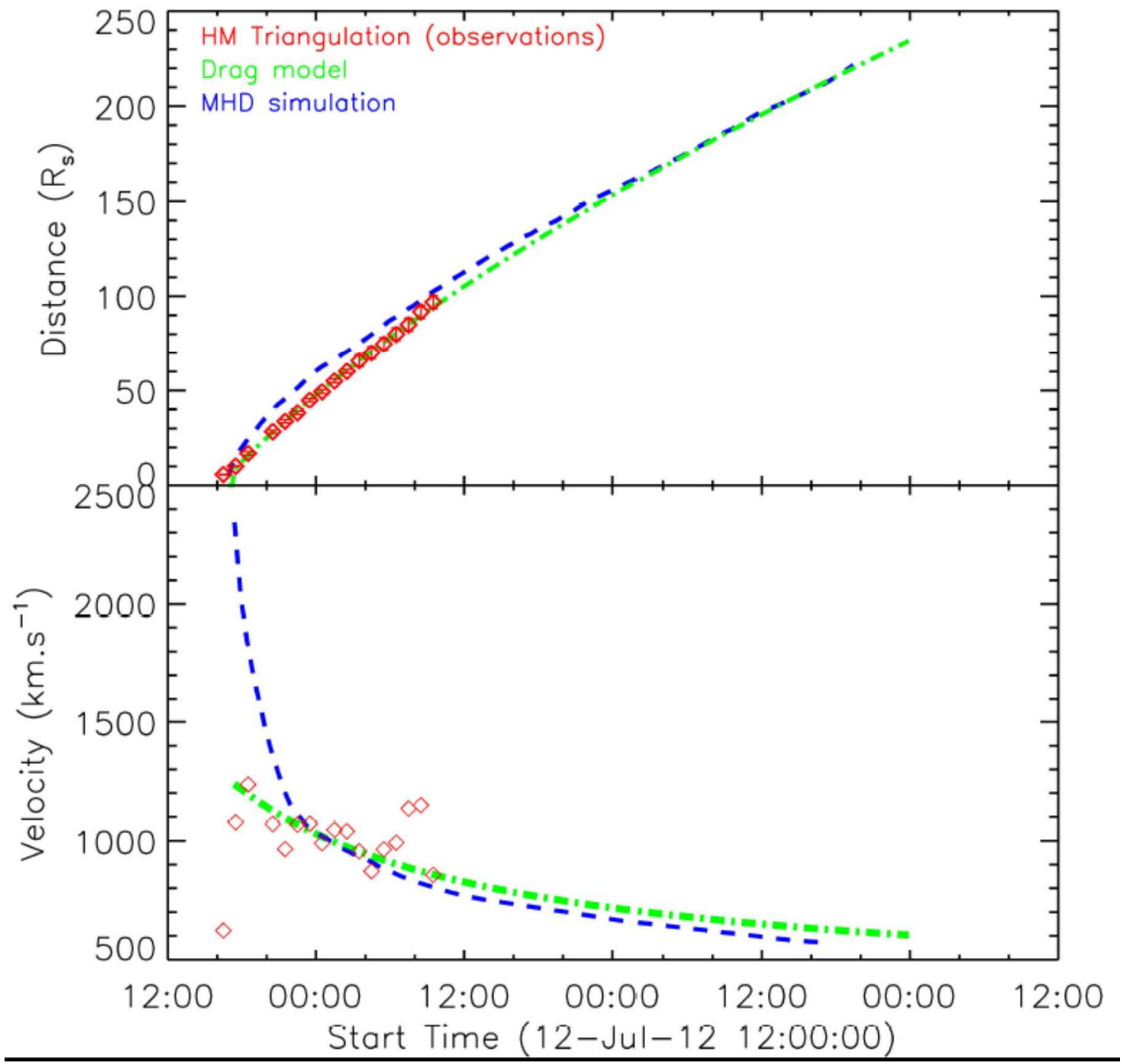

**Figure 11**. Time-height plot and the time-speed plot of the shock from near the Sun to the Earth. The blue dashed lines, the green dash-dotted lines and the red diamonds indicate the time-height and the time-speed distribution of the shock front from the simulation, from the shock drag model based on the observations and from the HM triangulation method based on the observations.

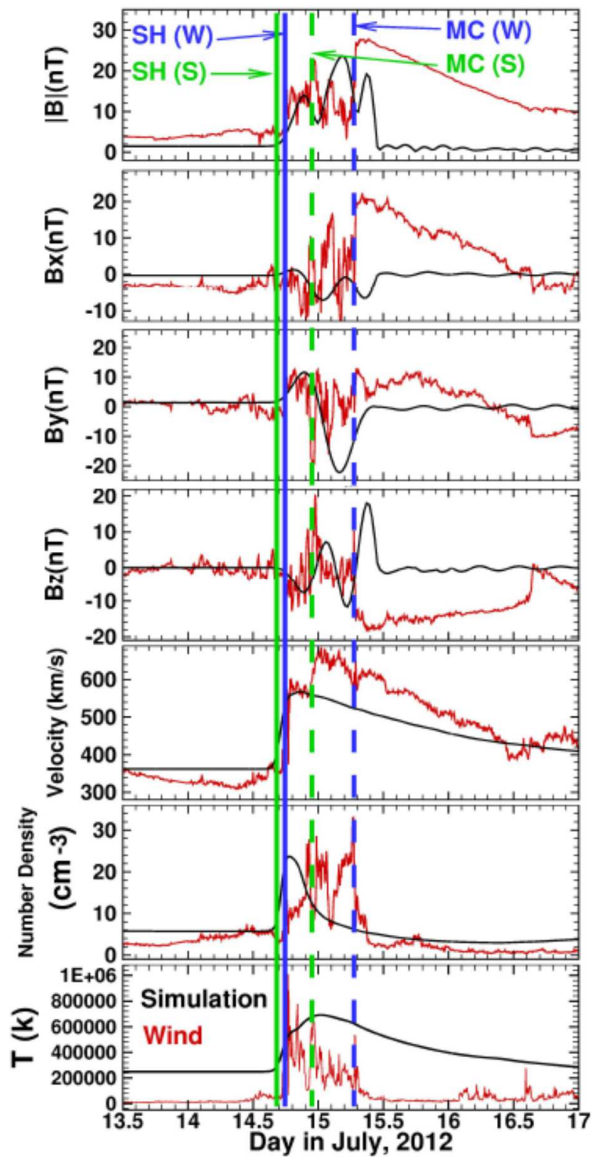

**Figure 12.** A comparison of the MHD simulation of the magnetic field and plasma parameters using the measured (Wind spacecraft) magnetic field and solar wind parameters at 1 AU. The black lines denote simulation parameters, and the red lines denote the measured parameters by Wind, (top to bottom) the magnetic field strength |B| (nT), $B_x$ (nT), $B_y$ (nT), $B_z$ (nT) at GSE coordinate system, the velocity (km/s), the proton density (cm$^{-3}$) and the proton temperature (K). The blue (or green) vertical solid lines

indicate the arrival time at 1AU of the shock and the blue (or green) vertical dashed lines denote the arrival time of the magnetic cloud, from Wind data (or Simulation).

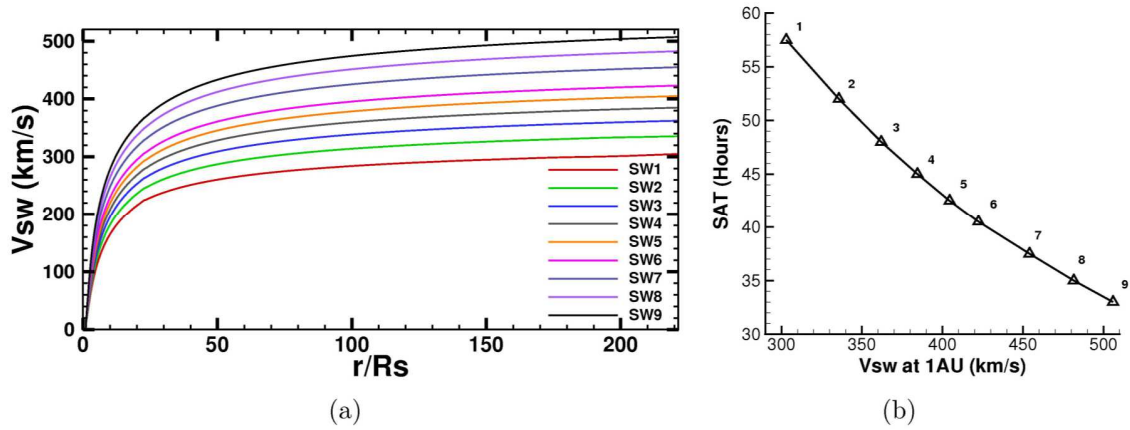

(a)  (b)

**Figure 13**. (a) Averaged background solar wind speed ($V_{sw}$) vs. height in nine cases which is presented from SW1 to SW9; (b) Shock Arrival Time (SAT) in the nines cases vs. $V_{sw}$ at 1AU.